\documentclass[letterpaper]{article} 
\usepackage[]{aaai23}  
\usepackage{times}  
\usepackage{helvet}  
\usepackage{courier}  
\usepackage[hyphens]{url}  
\usepackage{graphicx} 
\usepackage{natbib}  
\usepackage{caption} 
\usepackage{algorithm}
\usepackage{algorithmic}
\usepackage{multirow}
\usepackage{tabularray}
\usepackage{xcolor}
\usepackage{optidef}
\usepackage{tabularx}
\usepackage{adjustbox}
\usepackage{multirow}
\usepackage{float}
\usepackage{pifont}
\usepackage{multirow}
\usepackage{makecell}
\usepackage{graphicx}
\usepackage{dcolumn}
\usepackage{multirow}
\usepackage{array}
\usepackage{subcaption}
\usepackage{subfloat}
\usepackage{siunitx}
\usepackage{multirow}
\usepackage{hhline}
\usepackage{graphicx}
\usepackage{newfloat}
\usepackage{listings}
\usepackage{rotating}
\usepackage{hhline}
\usepackage{graphicx}
\usepackage{dcolumn}
\DeclareCaptionStyle{ruled}{labelfont=normalfont,labelsep=colon,strut=off} 
\floatstyle{ruled}
\newfloat{listing}{tb}{lst}{}
\floatname{listing}{Listing}

\usepackage{xcolor}

\title{Tight Sampling in Unbounded Networks}

\author{
    Kshitijaa Jaglan\equalcontrib \textsuperscript{\rm 1},
    Meher Chaitanya\equalcontrib \textsuperscript{\rm 2},
    Triansh Sharma \textsuperscript{\rm 1},
    Abhijeeth Singam \textsuperscript{\rm 1},
    Nidhi Goyal \textsuperscript{\rm 3},
    Ponnurangam Kumaraguru \textsuperscript{\rm 1},
    Ulrik Brandes \textsuperscript{\rm 2}
  }

 \affiliations {
     \textsuperscript{\rm 1} IIIT Hyderabad\\
     \textsuperscript{\rm 2} Social Networks Lab, ETH Zürich\\
     \textsuperscript{\rm 3} IIIT Delhi\\
      kshitijaa.jaglan@research.iiit.ac.in,
      mpindiprolu@ethz.ch, trianshsharma@gmail.com,  singam.abhijeeth@gmail.com, nidhig@iiitd.ac.in,  pk.guru@iiit.ac.in, ubrandes@ethz.ch
 }

\newcolumntype{d}[1]{D{,}{,}{#1}}

\begin{document}
\maketitle

\begin{abstract}
The default approach to deal with the enormous size and limited accessibility
of many Web and social media networks is to sample one or more subnetworks
from a conceptually unbounded unknown network.
Clearly, the extracted subnetworks will crucially depend on the sampling scheme.
Motivated by studies of homophily and opinion formation,
we propose a variant of snowball sampling 
designed to prioritize inclusion of entire cohesive communities
rather than any kind of representativeness, breadth, or depth of coverage.
The method is illustrated on a concrete example,
and experiments on synthetic networks suggest that it behaves as desired.
\end{abstract}

\section{Introduction}
\label{introduction}
Online social networks such as Twitter are a valuable source of information for research on various questions in the social sciences, not least because they contain vast amounts of process-generated data \cite{ANTONAKAKI2021114006}. However, obtaining full-size networks from platforms is impossible for researchers due to access limitations, and prohibitive due to volume. Sampling from massive online networks with millions or even billions of users thus presents a fundamental challenge for social network research~\cite{ruths2014studies}.

Common sampling schemes rely on one, or both, of two main techniques~\cite{ahmed2013network}: retrieval based on attributes such as demographics or tweet content (node-based or edge-based sampling), and seed-set expansion by following incoming or outgoing links (topology-based sampling) \cite{survey_of_twitter_methods, leskovec_FF_sampling}. 

Seed-set expansion is related to graph exploration and snowball sampling, where elements of a network of unknown size are discovered only through adjacency with already explored parts. If the underlying networks exhibit small-world characteristics, as many social networks do, the boundary of connectivity-based sampling methods quickly covers distant parts of the network.
When the research goal is study homophily and other social regularities, completeness of cohesive groups is a more important sampling criterion than coverage of the network. Our problem is, therefore, closely related to local clustering with seeds, and especially relevant in conceptually unbounded networks such as Twitter.

To this end, we propose a novel snowball-type sampling scheme that is designed to prioritize sampling within the cohesive subgroups or local clusters around a given set of seed nodes in the (multiplex) network.
The approach thus resembles seeded community detection, where the objective is to determine a locally dense subgraph containing a seed node or a set of seed nodes, except that in our case the graph is only partially known. 
Common clustering objectives such as low conductance or high modularity \cite{exponentially_twisted_sampling, zhang2018understanding, newman2006modularity} are difficult to optimize in such settings,
because the use of methods such as approximate PageRank~\cite{andersen2006local} or random-walk techniques \cite{spielman2004nearly} require large parts of the graph in which a seeded community resides to be available.
Without such information, and confronted with rapid expansion of the boundary around the sampled network, we prioritize the selection of nodes based on their likeliness to add to cohesive groups in the sample.

Our approach generalizes a technique known as maximum adjacency search~\cite{cai1993partitioning} that has been used prominently to find minimum graph cuts by repeatedly expanding from a seed node~\cite{stoer1997simple}. 
We replace the basic maximum-adjacency criterion with a generalized priority obtained from a combination of different forms of interaction in social media, such as likes, retweets, replies, and quotes with empirically calibrated weights. 
Specifically for sampling subnetworks on Twitter, we prioritize profiles outside the current sample set that show maximum levels of engagement with profiles inside. 
The evolution of sampled networks is demonstrated on empirical and synthetic data,
and we conclude that our method effectively prioritizes local clusters around seeds. 

In summary, our main contributions are as follows.
\begin{itemize}
    \item Application of a maximum-adjacency principle to snowball-sampling to expand seed sets while staying within local communities.
    \item Generalization of the maximum-adjacency criterion to weighted multiplex networks. Specifically for Twitter, we propose an empirically calibrated weighting scheme to combine types of interaction.
    \item Provision of a Twitter dataset focusing on the interactions within communities that engage with a publicly available set of influential profiles.
\end{itemize}

\section{Sampling Social Media Networks}
The majority of methods used for social media networks begin by sampling a set of profiles, links, or interactions and expand the network by traversing (parts of) their neighborhoods. In this section, we provide a brief overview of various relevant sampling approaches.

\paragraph{Node-based sampling methods:}
These can be as simple as the subgraph induced by a set of nodes sampled uniformly at random~\cite{ahmed2013network}. This technique is simple to approximate the direct properties of nodes, such as degree distribution, but it does not preserve the connectivity. More sophisticated node-based sampling approaches, such as the PageRank and PageRank-with-restarts methods, choose sampled nodes based on their PageRank scores and construct the induced subgraph on them~\cite{rozemberczki2020karate}. 

\paragraph{Traversal-based sampling methods:} Given one or more starting nodes (``seeds''), these are strategies adding more nodes by traversing the underlying graph from already sampled nodes. Some strategies use traditional graph traversals such as breadth-first and depth-first search~\citep{Giudice2019AlgorithmsFG}. In each iteration, a new node is chosen depending on its earliest (breadth-first) or latest (depth-first) discovery time. Generally, a breadth-first search can result in a denser cover and has been shown to be biased towards high-degree nodes~\cite{ye2010crawling}. The work by \citet{kurant2011towards} addressed this bias by suggesting analytical solutions to correct it. Snowball sampling is another traversal-based sampling strategy that maintains the network connectivity using the breadth-first approach but suffers from boundary bias where the peripheral nodes sampled in the final iteration have many missing neighbors~\cite{SB_sampling}. A large class of traversal-based sampling strategies are based on Random Walks (RW). RW sampling techniques commence a random walk (single or multi-dimensional) starting from seed nodes and construct a Markov chain by iteratively choosing a random neighbor~\cite{gjoka2010walking, ribeiro2010estimating, avrachenkov2010improving}. These techniques are inherently biased towards high-degree nodes \cite{hu2013survey}. Metropolis-Hastings random walk sampling strategies overcome this bias by making the random walker visit low-degree nodes~\cite{hubler2008metropolis, stutzbach2006unbiased, li2015random}. \citet{liu2019novel} incorporate a novel hybrid jump mechanism in Metropolis-Hastings random walk to avoid repetitive sampling within a small connected component. Forest Fire sampling, a hybrid of random walk-based methods and snowball sampling expands by burning a fraction of the outgoing links for each sampled node~\cite{leskovec_FF_sampling}. This fraction is drawn randomly from a geometric distribution with mean $\frac{p}{1-p}$ (the recommended value of $p$ is $0.7$, implying that, on average, each selected node burns $2.33$ neighbors). \citet{maiya2010sampling} proposed a community-preserving sampling approach by utilizing concepts from expander graphs to sample representative subgraphs that reflect the community structure of the original network by greedily constructing the sample with maximal expansion. Recently, \citet{zhang2023cluster} introduced expansion strategies for detecting clusters around seed nodes. These strategies involve including nodes in the sample through specific expansion techniques based on edge connectivity. However, all these approaches require large parts of the graph surrounding the seeded nodes to be available. In the next section, we provide a methodology to overcome the uncertainty of the unknown or unboundedness of the network to make the sampler stay within cohesive subgroups surrounding seeds.

\section{Tight Sampling}\label{sec:Methodology}

Our goal is to sample subgraphs of social media networks in such a way
that cohesive communities are covered in larger parts
before expanding further into the underlying network.
We refer to this as \emph{tight} sampling.
Since the network is assumed to be much larger than the targeted sample size,
say, all of Twitter, we think of it as unbounded.

Formally, we assume the existence of
an infinite, initially unknown directed graph $G=(V,E)$
representing a vast social media network.
Edges represent social relations between members of the network
and will be described more concretely below,
where we also introduce edge weights. 
We further assume that knowing a vertex~$v\in V$,
we can also obtain the set $N^-(v)=\{u\in V: (u,v)\in E\}$ of in-neighbors
with edges directed to~$v$;
the set of out-neighbors $N^+(v)$ is defined symetrically.

Given a finite set $V_s\subset V$ of \emph{seed} vertices,
we want to extract a subgraph $G[S]$
induced by a finite set of sample vertices $S\subset V$
that includes the seeds, $V_s\subseteq S$.
Starting from the seeds, vertices are sampled one at a time,
and each newly sampled vertex must be an in-neighbor of a vertex sampled earlier.
In other words, we aim for a sampling strategy
that traverses edges backwards.
Thus, we successively add vertices that relate to those already included.

For notational simplicity, we omit timestamps
and refer to the set of currently sampled vertices, or \emph{insiders}, as~$S$.
Candidate vertices that may be sampled next
are all in-neighbors $N^-(S)=\bigcup_{v\in S} N^-(v)$ not yet in~$S$.
We refer to the vertices in $N^-(S)\setminus S$ as \emph{outsiders}.

The \emph{boundary}~$\partial(S)$ of a current sample~$S$
is the set of all edges directed from outsiders to insiders,
i.e., the edges crossing a \emph{directed cut}.
Since our objective is to keep this boundary small,
we sample outsiders that have the maximum number of edges directed to insiders.
This is a directed version of maximum-adjacency search,
and greedily removes edges from the boundary. 
Note that we do not know the in-neigborhood of a vertex prior to its sampling,
so that we can not make any guarantees whether the new boundary is smallest possible.

In summary, we sample a vertex-induced subgraph
by expanding a set of seed vertices one vertex at a time,
where the vertex selected is the outsider with
the largest number of edges directed to insiders,
i.e., by maximum-adjacency search.
In the next section, we extend this principle to weighted graphs
that integrate multiple types of relations and interactions
in social media networks, and then validate the outcome.

\section{Weighted Edges from Multiplex Relations}

Social media typically combine multiple types of
relations such as friending or following with
interactions such as liking or forwarding.
In order to sample subgraphs
in which the most cohesively related groups are relatively intact,
we propose an empirically calibrated aggregation
into a single weighted relation.
This will allow for straightforward generalization of
the maximum-adjacency principle
from counting edges to the sum of their weights.

As detailed in the following three subsections, weights are computed
by deciding first on the patterns of interaction to distinguish, 
and then combining their re-scaled frequencies of occurrence.

\subsection{Interaction patterns} 

Because of our specific interest in social influence on Twitter, 
we consider four kinds of relations as indicators of engagement
with information shared by other users via tweets:
likes, retweets, replies, and quotes.
First, the interaction pattern of a user~$i$
with a tweet~$t$ authored by~$i$
is represented by the characteristic vector
$I_t(i, j) = x \in X$ of interaction types, $x \in X = \{0,1\}^4$.
Here, binary values $\{0,1\}$ denote the presence or absence
of a particular form of engagement
from the set \{\textit{like, retweet, reply, quote}\}.
For example, if a user~$j$ retweets and quotes a tweet~$t$ of user~$i$,
there is a directed edge from~$j$ to~$i$
labeled with interaction pattern $I_t(i,j)=0101$.
We omit indices~$i$ and~$j$ if they is clear from the context.
Note that for a single tweet and interacting user
we only consider the presence or absence of forms of engagement,
not the number of their respective instances.

\subsection{Frequency of occurrence}

When counting interaction patterns 
it is sometimes desirable to count occurrences of one pattern 
also toward the frequency of another,
because it may or may not matter
whether additional types of interaction are present. 
We distinguish three cases.

    \paragraph{Distinct interaction patterns.} 
A pair of a tweet and interacting user contributes
to the frequency of an interaction pattern
only if the user engages with the tweet in exactly this pattern.
A user's engagement is counted as an occurrence of pattern $x=1100$,
for instance,
if and only if the user \emph{likes} and \emph{retweets}
and does not \emph{reply} or \emph{quote}.

    \paragraph{Nested interaction patterns.} 
A pair of a tweet and interacting user contributes
to the frequency of an interaction pattern
if the user engages with the tweet including this pattern.
A user's engagement is counted as an occurrence of pattern $x=1100$,
for instance,
if and only if the user \emph{likes} and \emph{retweets}
and does or does not \emph{reply} or \emph{quote}.

    \paragraph{Audience-facing interactions (A-F).}
We posit that likes and replies are more personal forms of interaction
and usually directed at the author of a tweet,
whereas retweets and quotes tend to be aimed at visibility
by signaling an interaction to followers.
We therefore introduce a third method of counting by treating 
retweets and quotes as interchangable types of interaction.
We thus have $X=\{001,010,011,100,101,110,111\}$,
reducing the effective number of patterns from~15 to seven.
Merging of retweets and quotes has been applied in other studies
for instance on the Higgs Boson Twitter dataset \cite{de2013anatomy}.

\subsection{Importance scaling} 

Interaction types occur at different rates
and therefore potentially signal different levels of engagement. 
Liking is the most frequent form of interaction,
but is therefore assumed to be less informative than, say, quoting.
To determine the relative importance of interaction patterns,
we therefore first assess their empirical prevalence
and then assign a weight inversely proportional to it.

Assume we are given an empirical sample~$S$ of insiders as well as
their tweets, and the interactions with them.
Denote by $T\supseteq S$ the set of interacting users.
Furthermore let $n(i,x,j)$ denote the number of times
that any user~$j\in T$ engaged with the tweets
of a user~$i\in S$ using interaction pattern~$x\in X$,
and let $N=\sum_{i,x,j} n(i,x,j)$ be the overall number of pattern occurrences.
Recall that, say,
multiple replies of the same user to the same tweet are counted only once.
To derive importance weights for the types of interaction, 
we first distinguish three approaches to normalizing frequencies.

	\paragraph{Global normalization.}
Ignoring the users involved, the overall frequency of interaction pattern~$x\in X$
is given by $n(x)=\sum_{i\in S,j\in T} n(i,x,j)$.
The relative frequency of pattern~$x$, normalized globally, is then defined as
$$
  \eta(x) = \frac{n(x)}{N}~.
$$

	\paragraph{Source normalization.}
Users spreading information
may see very different patterns of engagement with their tweets.
An alternative approach is therefore to normalize interaction patterns
by the average engagement that sources of information receive,
$$
  \overleftarrow{\eta}(x)
	= \frac{1}{|S|}\sum_{i\in S}\frac{n(i,x)}{\overleftarrow{N}(i)}
$$
where $n(i,x)=\sum\limits_{j\in T} n(i,x,t)$
and $\overleftarrow{N}(i)=\sum\limits_{x\in X} n(i,x)$.

	\paragraph{Target normalization.}
Symmetrically, users interacting with information published by others
may exhibit very different patterns of engagement.
An alternative approach is therefore to normalize interaction patterns
by the average engagement that consumers of information display,
$$
  \overrightarrow{\eta}(x)
	= \frac{1}{|T|}\sum_{j\in T}\frac{n(x,j)}{\overrightarrow{N}(j)}
$$
where $n(x,j)=\sum\limits_{i\in S} n(i,x,t)$
and $\overrightarrow{N}(j)=\sum\limits_{x\in X} n(x,j)$.

\bigskip
For the purpose of this paper we are balancing all of the above three perspectives
by determining a distribution that minimizes the sum of squared errors
with respect to the alternatives,
i.e., we find a non-negative vector $\eta^*(x)$ for the set of patterns~$X$ such that
$$
  \sum_{x\in X} (\eta^*(x)-\eta(x))^2
              + (\eta^*(x)-\overleftarrow{\eta}(x))^2
              + (\eta^*(x)-\overrightarrow{\eta}(x))^2
$$
is minimum.
In the spirit of Horvitz-Thompson importance sampling,
we finally determine influence weights $\omega(x)$ for the interaction patterns
as the inverse of their balanced normalized frequencies,
$$
  \omega(x) = \frac{1}{\eta^*(x)}
$$
In practice, we use entries $\omega^*(x)$ rounded to two decimals
for simplicity and robustness.
\begin{table*}[h]
\begin{subtable}{1\linewidth}
\centering
    \resizebox{\textwidth}{!}{
    \begin{tabular}{|l|*{12}{S[table-format=4.4]|}S[table-format=4.4]|}
\hline
\multicolumn{1}{|c|}{} & \multicolumn{2}{c|}{$\eta(x)$} & \multicolumn{2}{c|}{$\overleftarrow{\eta}(x)$} & \multicolumn{2}{c|}{$\overrightarrow{\eta}(x)$} & \multicolumn{2}{c|}{$\eta^*(x)$} & \multicolumn{2}{c|}{$ \omega(x)$} & \multicolumn{2}{c|}{\begin{tabular}[c]{@{}c@{}}$\omega^*(x)$\\ \end{tabular}} \\
Interaction type & \multicolumn{1}{c}{Distinct} & \multicolumn{1}{c|}{Nested} & \multicolumn{1}{c}{Distinct} & \multicolumn{1}{c|}{Nested} & \multicolumn{1}{c}{Distinct} & \multicolumn{1}{c|}{Nested} & \multicolumn{1}{c}{Distinct} & \multicolumn{1}{c|}{Nested} & \multicolumn{1}{c}{Distinct} & \multicolumn{1}{c|}{Nested} & \multicolumn{1}{c}{Distinct} & \multicolumn{1}{c|}{Nested} \\ 
\hhline{|=============|}
0001 & 2.2560 & 2.8947 & 2.0575 & 2.9979 & 1.4276 & 2.1090 & 1.9137 & 2.6672 & 0.5225 & 0.3749 & \emph{0.52} & \emph{0.37} \\
0010 & 7.9125 & 9.4169 & 5.5468 & 7.9285 & 5.9666 & 7.9333 & 6.4753 & 8.4263 & 0.1544 & 0.1186 & 0.15 & 0.12 \\
0011 & 0.3272 & 0.0583 & 0.0645 & 0.1020 & 0.0367 & 0.0793 & 0.0047 & 0.0799 & 22.3920 & 12.5190 & 22.4 & 12.52 \\
0100 & 6.0684 & 15.9760 & 6.8172 & 18.4080 & 6.2281 & 18.4700 & 6.3712 & 17.6180 & 0.1569 & 0.0568 & 0.16 & 0.06 \\
0101 & 0.0687 & 0.2199 & 0.1092 & 0.3414 & 0.0864 & 0.2801 & 0.0881 & 0.2805 & 11.3490 & 3.5652 & 11.35 & 3.6 \\
0110 & 0.0860 & 0.3692 & 0.1641 & 0.5828 & 0.1015 & 0.5901 & 0.1172 & 0.5141 & 8.5331 & 1.9452 & 8.53 & 1.95 \\
0111 & 0.0018 & 0.0102 & 0.0034 & 0.0203 & 0.0031 & 0.0238 & 0.0028 & 0.0181 & 360.1600 & 55.1890 & \textbf{360.1} & 55.2 \\
1000 & 72.3500 & 83.5740 & 71.6710 & 85.2370 & 72.4130 & 86.1500 & 72.1440 & 84.9870 & 0.0139 & 0.0117 & 0.014 & 0.01 \\
1001 & 0.3707 & 0.5355 & 0.5174 & 0.7633 & 0.3457 & 0.5551 & 0.4113 & 0.6180 & 2.4314 & 1.6182 & 2.43 & 1.6 \\
1010 & 1.0871 & 1.3840 & 1.7172 & 2.1497 & 1.3212 & 1.8254 & 1.3752 & 1.7864 & 0.7272 & 0.5598 & 0.73 & 0.6 \\
1011 & 0.0154 & 0.0238 & 0.0172 & 0.0341 & 0.0188 & 0.0395 & 0.0171 & 0.0324 & 58.4740 & 30.8210 & 58.5 & 30.8 \\
1100 & 9.3286 & 9.7510 & 10.6870 & 11.3150 & 11.3950 & 12.0510 & 10.4700 & 11.0390 & 0.0955 & 0.0906 & 0.095 & 0.09 \\
1101 & 0.1410 & 0.1495 & 0.2118 & 0.2287 & 0.1699 & 0.1906 & 0.1743 & 0.1896 & 5.7386 & 5.2742 & 5.74 & 5.3 \\
1110 & 0.2730 & 0.2815 & 0.3984 & 0.4153 & 0.4648 & 0.4855 & 0.3787 & 0.3941 & 2.6402 & 2.5375 & 2.64 & 2.5 \\
1111 & 0.0084 & 0.0084 & 0.0169 & 0.0169 & 0.0207 & 0.0207 & 0.0153 & 0.0153 & 65.1750 & 65.1750 & 65.2 & \textbf{65.2} \\
\hline
\end{tabular}}
\caption{Weights per interaction pattern involving \textit{like, retweet, reply, quote} for distinct and nested interactions.}
\vspace{0.5cm}
    \begin{tabular}{|l|*{7}{S[table-format=4.4]|}S[table-format=4.4]|}
    \hline
     & {$\eta(x)$} & {$\overleftarrow{\eta}(x)$} & {$\overrightarrow{\eta}(x)$} & {$\eta^*(x)$} & {$\omega(x)$} & {$\omega^*(x)$}\\ 
    \hhline{|=======|}
    001 & 19.0910 & 21.7480 & 20.8590 & 20.5660 & 0.0486 & 0.05 \\
    010 & 9.4170 & 7.9285 & 7.9333 & 8.4267 & 0.1187 & 0.12 \\
    100 & 83.5740 & 85.2370 & 86.1500 & 84.9870 & 0.0117 & 0.01 \\
    101 & 10.4360 & 12.3060 & 12.7960 & 11.8460 & 0.0844 & 0.08 \\
    110 & 1.3840 & 2.1496 & 1.8254 & 1.7864 & 0.5599 & 0.6 \\
    111 & 0.3137 & 0.4663 & 0.5456 & 0.4419 & 2.2624 & 2.3 \\
    \hline
    \end{tabular}
\caption{Weights per interaction pattern involving \textit{like, retweet, reply, quote} for  audience-facing interactions.}
\end{subtable}
\caption{Weights per interaction pattern involving distinct, nested, and audience-facing interactions.}
\label{tab:weights}
\end{table*}
\begin{figure*}[h!]
\centering
\begin{tabular}{cc}
\includegraphics[width=0.45\linewidth]{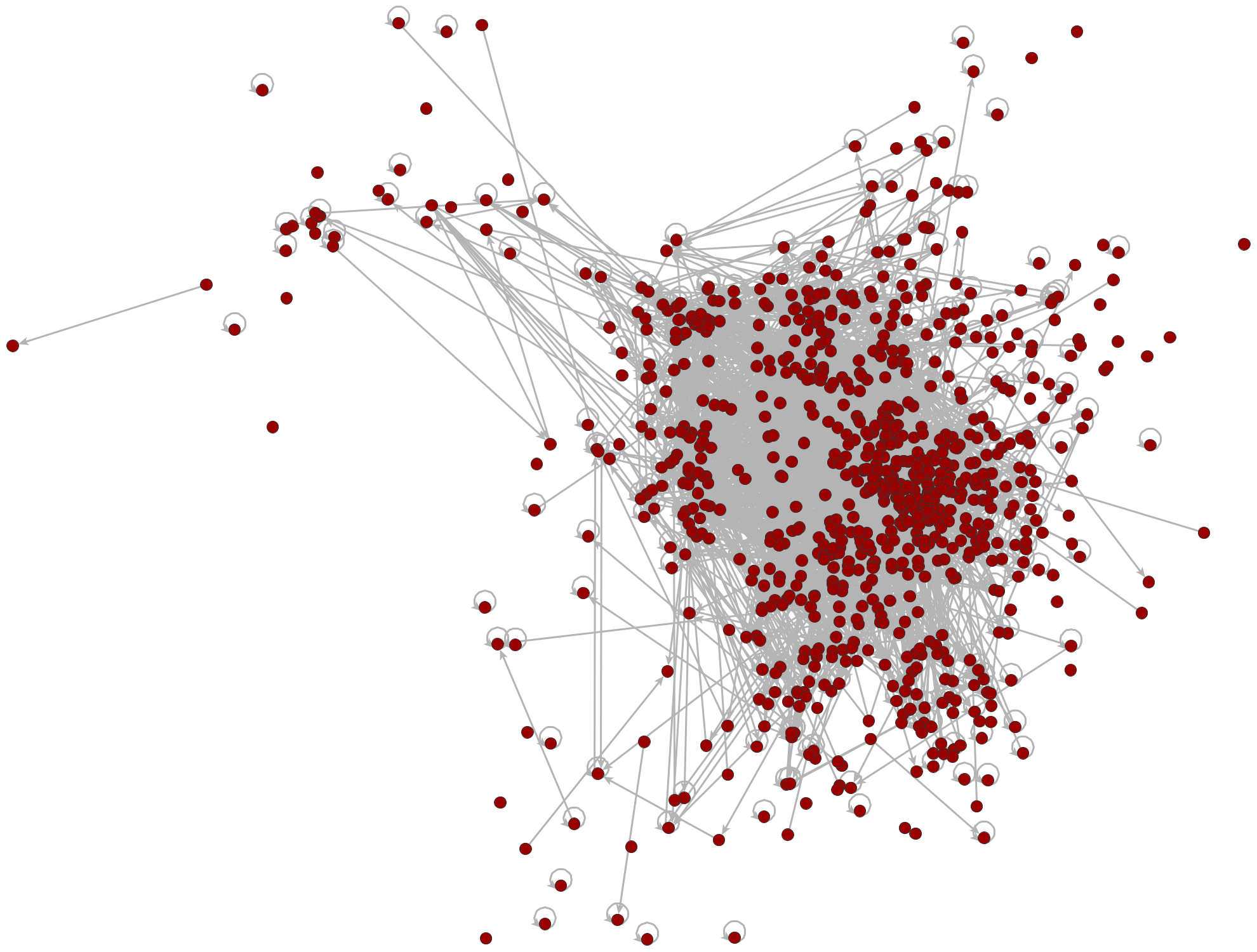} &
\includegraphics[width=0.45\linewidth]{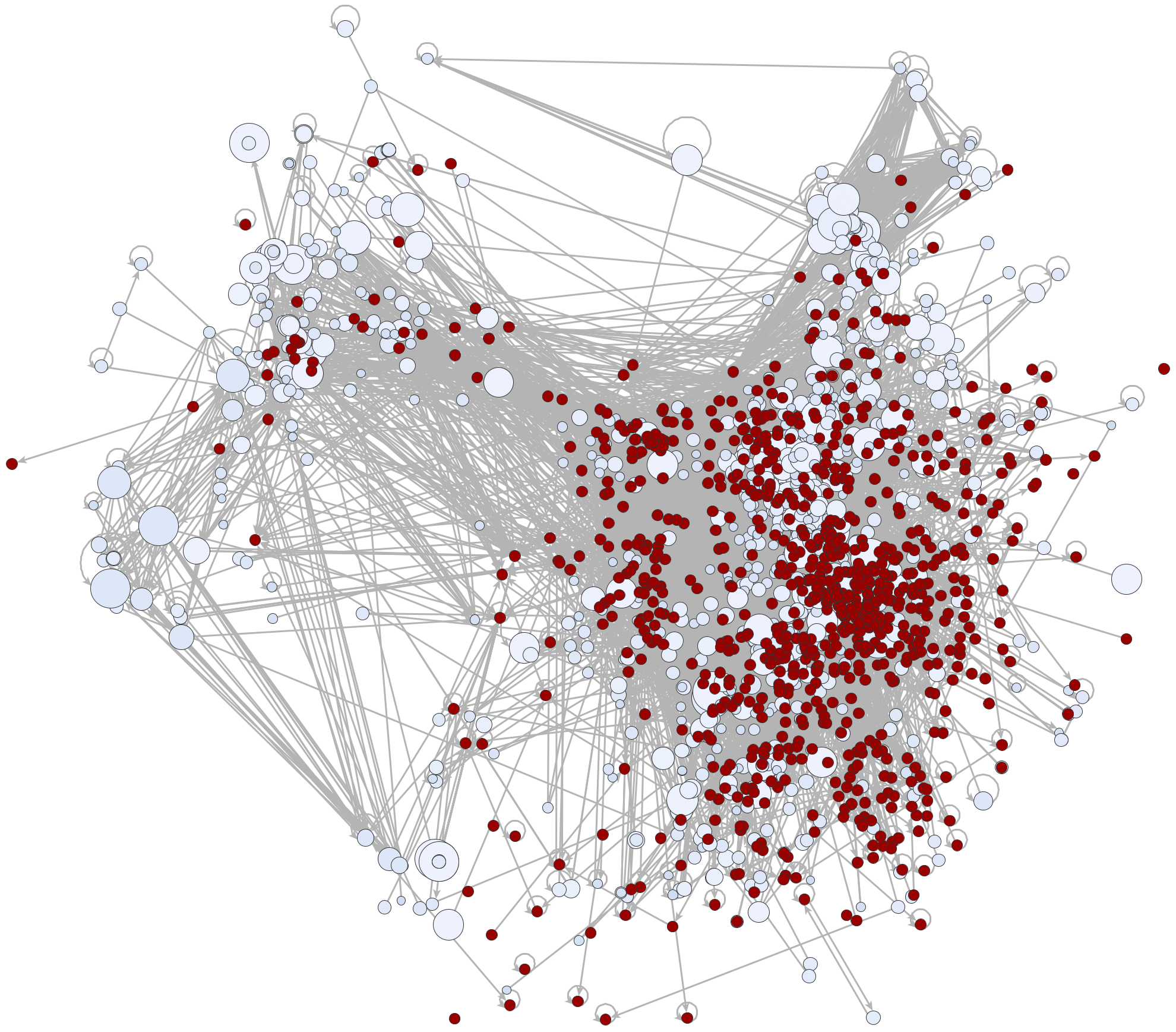} \\
(a) Seed nodes (DISMISS data) & (b) after 1000 timesteps\\[2ex]
\includegraphics[width=0.45\linewidth]{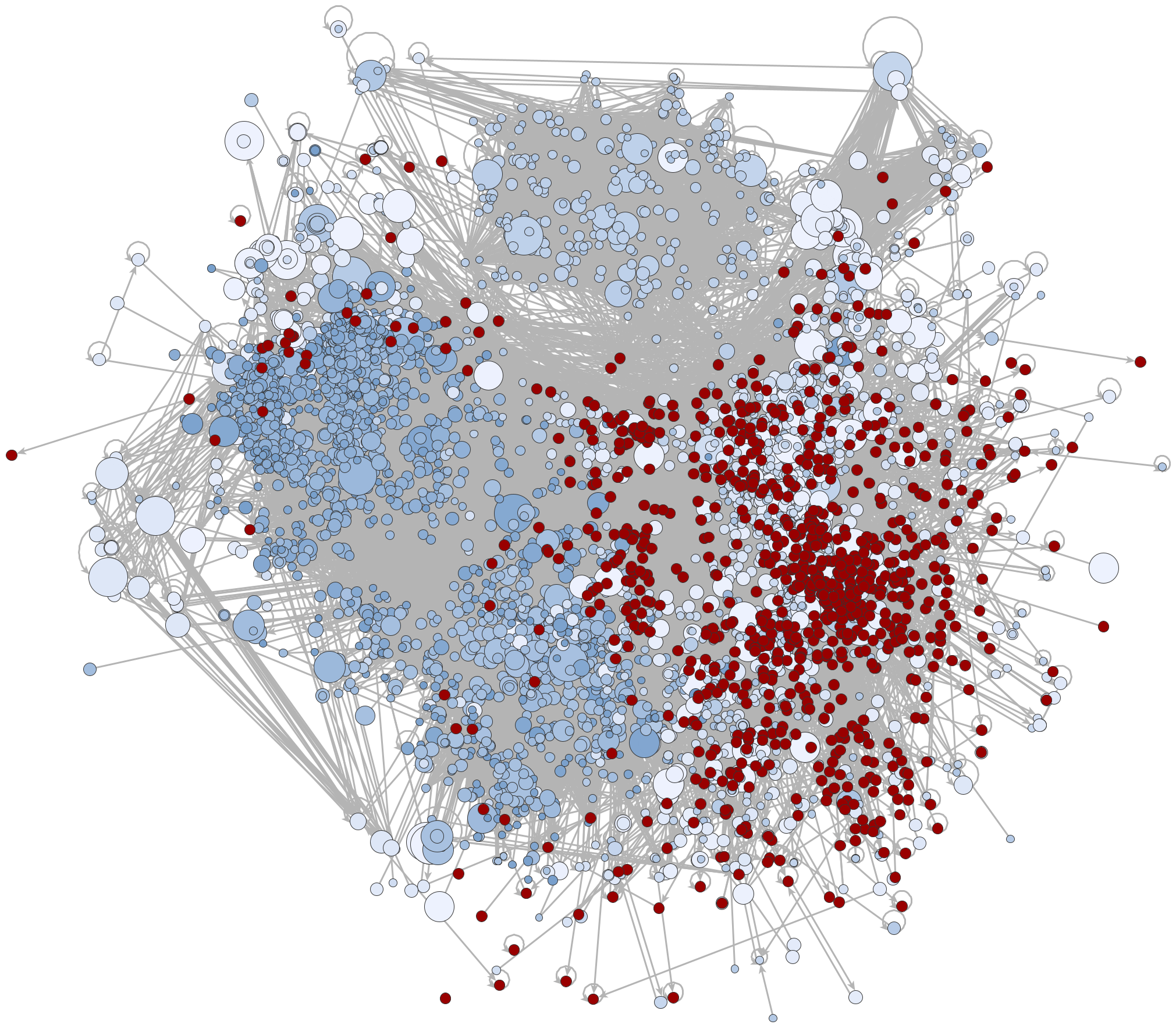} & \includegraphics[width=0.45\linewidth]{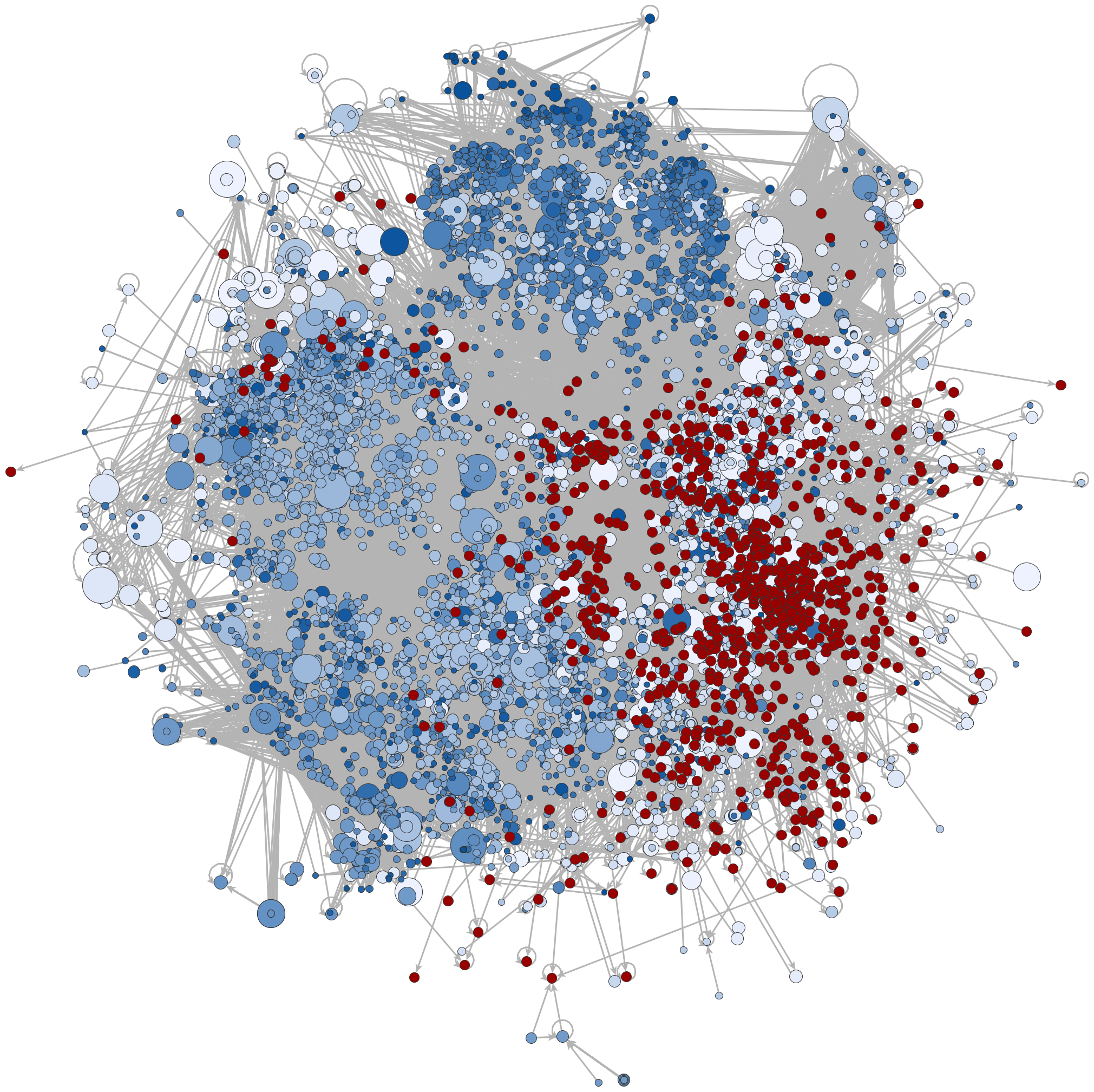} \\
 (c) After 4000 timesteps &(d) After termination (8000 timesteps) \\
\end{tabular}
\caption{Growth of the Twitter network during the expansion of the DISMISS data
by sampling nodes with maximum priority according to distinct interaction patterns. 
Each timestep corresponds to the move of an outsider into the insider set.
The layout is determined by the final network and constant across subfigures. 
Seed nodes are in red, and other nodes vary from light to dark blue 
based on the time of sampling.
Non-seed node sizes reflect their priority at the time of sampling.
While nodes added early are strongly interacting with the seed set,
spillover into other cohesive communities is readily observed.}
\label{fig:disjoint_network}
\end{figure*}

\section{Experiments}
We evaluate our sampling strategy by first creating synthetic networks in a controlled setting using stochastic blockmodels (SBM) \cite{holland1983stochastic}. This allows us to generate networks with predefined communities for which we can monitor how they are sampled. Following this controlled assessment, we proceed to evaluate our sampling approach in an empirical context by expanding a seed set on Twitter into a directed weighted network. Our findings on both, synthetic data and the empirical network, indicate that our sampling strategy improves on the coverage of cohesive communities when contrasted with random-based sampling approaches.
 
\subsection{Synthetic data}
In this section, we will describe the process of generating synthetic network data with planted communities. These networks constitute the simplest meaningful situations, in which the evolution of our sample can be observed most clearly.

\paragraph{Instances.}
We explore different networks  generated using SBM by varying block sizes, inter/intra block densities, and seed node distributions. Specifically, we explore three distinct block size settings: (1)  four blocks of sizes \{400, 800, 1200, 1600\}, (2) four blocks of sizes \{800, 1200, 1600, 2000\}, and (3) eight blocks of 1000 nodes each. For these three configurations, we derive the block probability matrix with consistent average degrees within each block $(\langle k' \rangle )$ and a uniform ratio of intra-block to inter-block edges, denoted as $r$, across all blocks. In this context, we define several key parameters: $n$ represents the total number of nodes in the SBM, $\rho_{ij}$ signifies the inter-cluster probability between block $i$ and block $j$, $n_i$ denotes the size of the $i^{th}$ block, $m_{ii}$ represents the number of edges within block $i$, $m_{i,*}$ indicates the total count of edges between block $i$ and all other blocks, and $\rho_{i,*}$ denotes the approximate density between block $i$ and the other blocks. Given a specific configuration characterized by $\langle k' \rangle$, $r$, and $b$ blocks with specified sizes, we derive the block probability matrix $P$ as follows:

In the case of the diagonal elements of the matrix $P$, the value of $\rho_{ii}$ is straightforwardly calculated as
$$\rho_{ii} = \frac{\langle k' \rangle}{n_i-1}$$

However, for non-diagonal elements, we determine the value of $\rho_{ij}$ using the ratio $r$, which represents the proportion of intra-block to inter-block edges as follows. 

\begin{align*}
    m_{ii} &= r \cdot m_{i,*} \\
    \frac{n_i \cdot \langle k'\rangle}{2} &= r \cdot \rho_{i,*} \cdot n_i \cdot (n-n_i) \\
    \rho_{i,*} &= \frac{\langle k' \rangle}{2 \cdot r \cdot (n-n_i)}    
\end{align*}

In the case of a block $(i,j)$ where $i \neq j$, since the previously calculated value is non-symmetric, we derive the final value for the respective cell by averaging $\rho$ values with respect to both the row and column:

\begin{equation}
\rho_{ij} = \frac{\rho_{i,*} + \rho_{j,*}}{2}
\end{equation}
where $\rho_{ij}$ represents the value in block $(i,j)$ for the block transition matrix $P$.

For our study, we have chosen the following values for $r$: {$\frac{1}{b-1}$, $0.5$, $1$, $2$, $4$, $8$}. These values have been carefully selected to facilitate an evaluation of the sampler's performance across a spectrum of community structure definitions. The minimum value of $r$ corresponds to a scenario in which, for each edge within block $i$, there are approximately $b-1$ edges connecting block $i$ to the other blocks. In this particular scenario, we observe a lack of distinct community structure, representing a case where our sampler struggles to identify clear community boundaries. By varying the values of $r$, we can effectively demonstrate the gradual changes in the sampler's performance.
 
It is important to highlight that while the intra-block average degree is fixed at $\langle k'\rangle$, the average degree of the entire network can vary due to the presence of inter-block edges (determined by $r$). Nevertheless, the process maintains uniform degree distributions across all blocks, ensuring that the sampler's preferences are not influenced solely by the presence of higher or lower degree nodes in specific communities. In our study, we set the value of $\langle k' \rangle$ to 10 for all SBM configurations.

\paragraph{Selection of seeds.}
In the case of the first two network configurations, which consist of four blocks each, we conducted experiments involving varying numbers of seed nodes per block. Specifically, we selected $20$ nodes per block from two blocks at a time, totaling $6$ possible combinations. We repeated this process with 50 nodes per block, resulting in a total of $12$ possible combinations. Furthermore, we explored the scenario in which each community was planted  with $20$ seed nodes, and similarly, we conducted experiments with $50$ nodes per block.

We conducted experiments with different seed node configurations in the network comprising eight communities, each containing 1000 nodes. These configurations included $[1]*8$, $[10]*8$, $[20]*2$, and $[20]*3$, where the notation $[i]*j$ indicates that there are $i$ seed nodes in each of the $j$ blocks, with the remaining blocks having zero seed nodes. We utilize random sampling to obtain the requisite number of nodes per block for all the aforementioned seed node configurations. Alternatively, we considered selecting nodes based on their degree centrality, both low and high, but throughout our experiments, we did not observe any significant disparities in the results.

\subsubsection{Sampling: }
For a given synthetic defined by its $P$ matrix and selection of seed users, we sample new nodes by employing the following expansion based strategies:

\begin{enumerate}
    \item \textbf{Maximum Adjacency Search (MAS)}. This strategy selects an outsider (non-seed node) with the  highest number of edges incident to the insider set.
    \item \textbf{Random Insider and MAS (RI\_MAS).} This strategy randomly selects an insider, $i$, and selects an outsider incident to $i$ based on  maximum adjacency search.
    \item \textbf{Random Outsider (RO).} This strategy randomly samples an outsider with uniform probability from the set of outsiders.
    \item \textbf{Random Insider and Random Outsider (RI\_RO).} We randomly select an insider followed by a random outsider incident to this insider.
\end{enumerate}

\subsubsection{Evaluation:}
\begin{figure}[h!]
\centering
\begin{subfigure}[b]{\linewidth}
    {{\includegraphics[width=\textwidth]{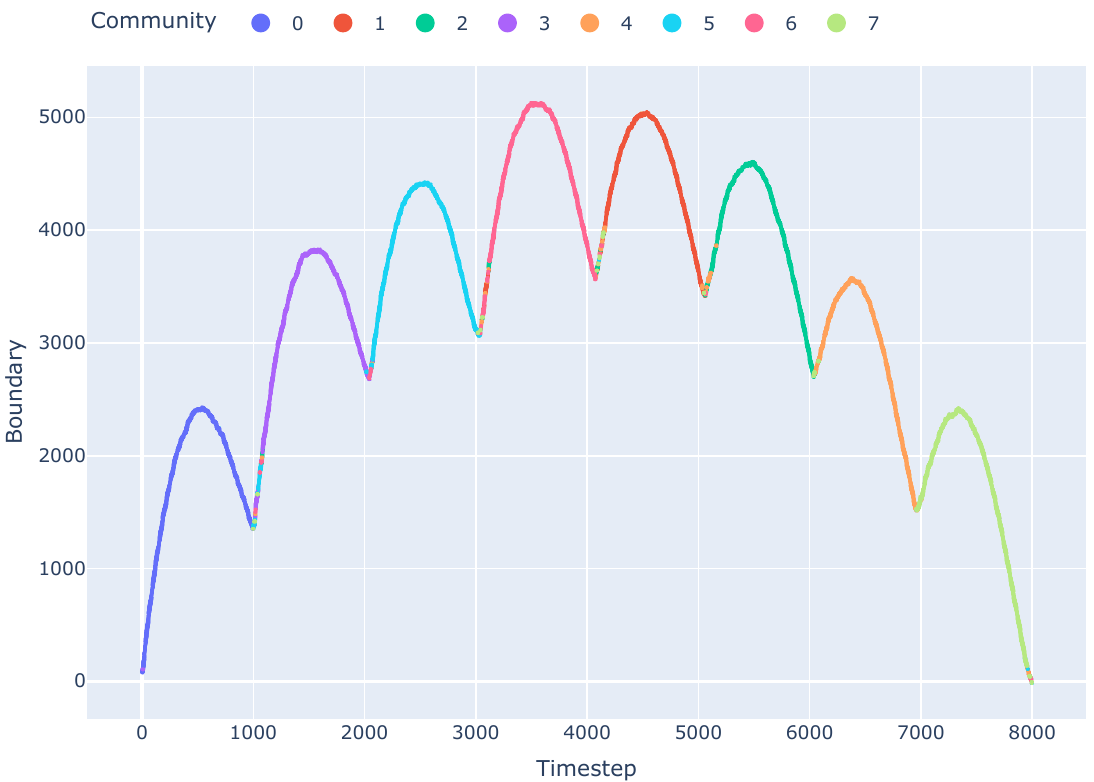}}}
    \caption{Evolution of network boundary with time}
\end{subfigure}
\begin{subfigure}[b]{\linewidth}
    {{\includegraphics[width=\textwidth]{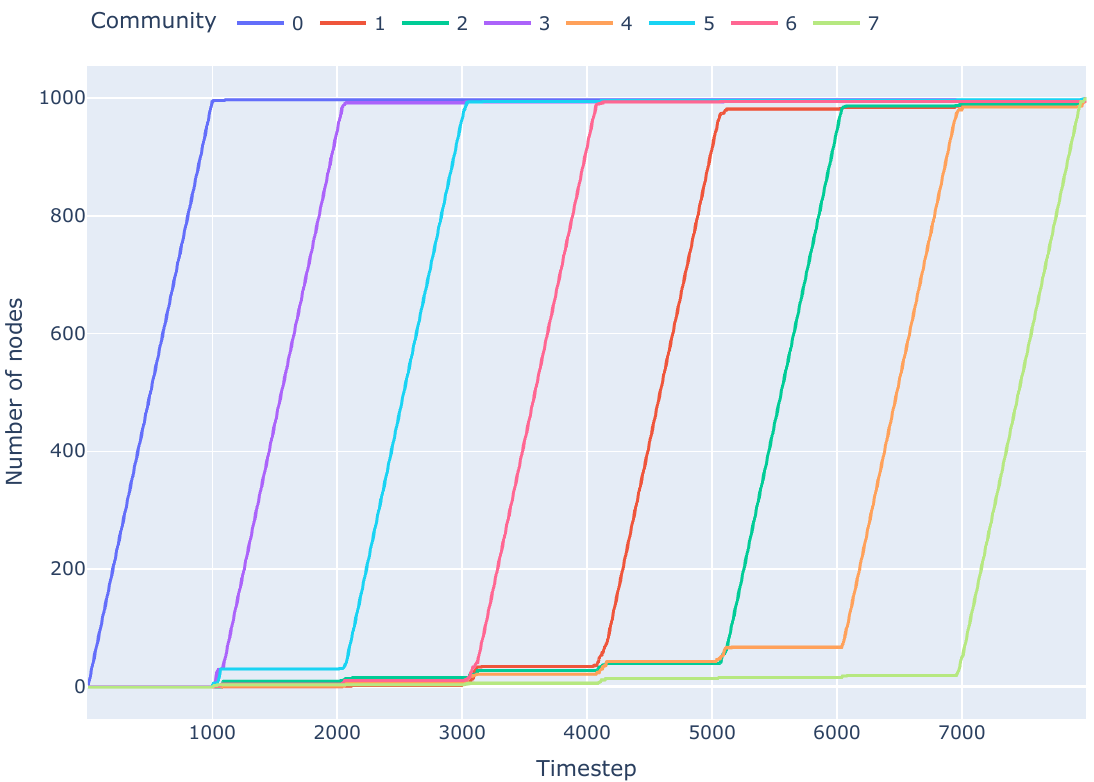}}}
    \caption{Community size evolution}
    
\end{subfigure}

\caption{Sampling using MAS for a network of 8 blocks with 1000 nodes and one seed node each. The ratio of intra cluster to inter cluster edges ($r$) and average degree within the block ($\langle k' \rangle$) are set as 4 and 10 respectively}
 \label{fig:sbm_1k_1_4_comm_evol_boundary}
\end{figure}

Our primary focus is directed towards the \emph{boundary vs. timestep} plot to assess the synthetic networks we have constructed in light of our objective. Additionally, we make use of community size evolution plots to discern which community is being sampled at a given time. As an illustrative example of the outcomes we aim to achieve with our sampling scheme, consider Figure \ref{fig:sbm_1k_1_4_comm_evol_boundary}(a), which showcases the boundary's dynamic changes in one of the synthetic network configurations.

In Figure \ref{fig:sbm_1k_1_4_comm_evol_boundary}(a), notable inflection points are clearly visible around timesteps 1000, 2000, 3000, etc., along with the commencement of a corresponding steep increase in the size of one of the communities as depicted in Figure \ref{fig:sbm_1k_1_4_comm_evol_boundary}(b). Here, an inflection point refers to the timestep at which the sampler starts sampling a new community.

\begin{figure}[h!]
\centering
\begin{subfigure}[b]{\linewidth}
    {{\includegraphics[width=\textwidth]{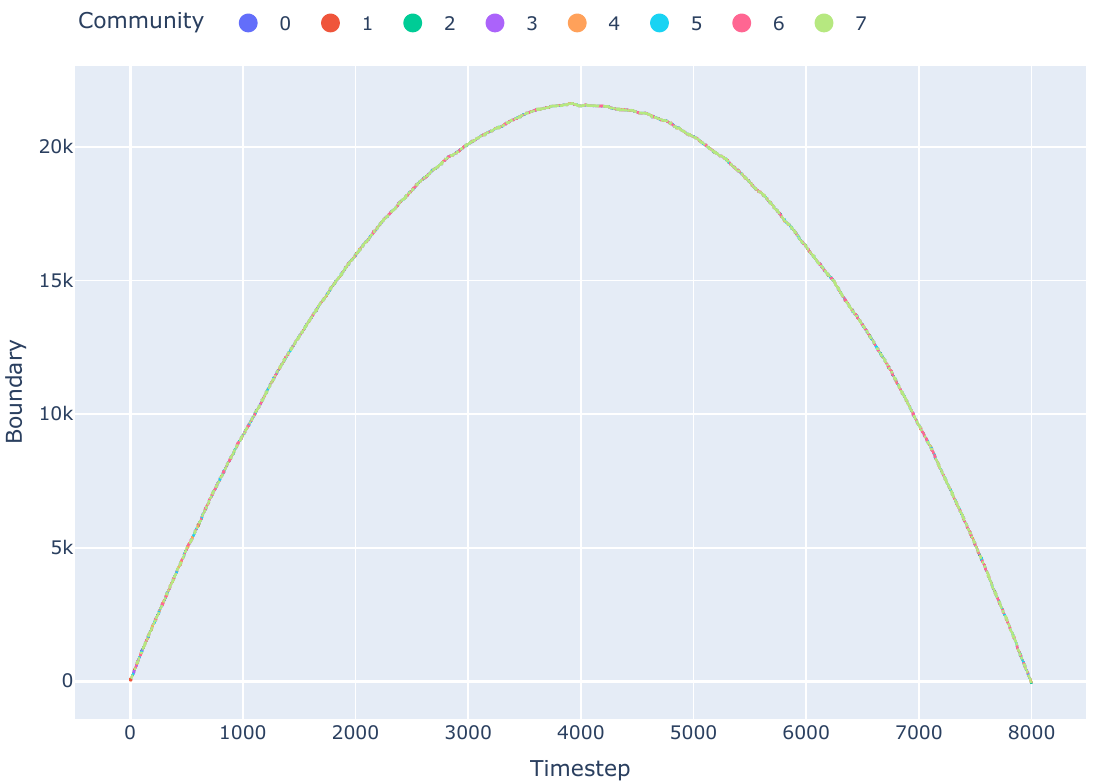}}}
    \caption{Evolution of network boundary with time}
\end{subfigure}
\begin{subfigure}[b]{\linewidth}
    {{\includegraphics[width=\textwidth]{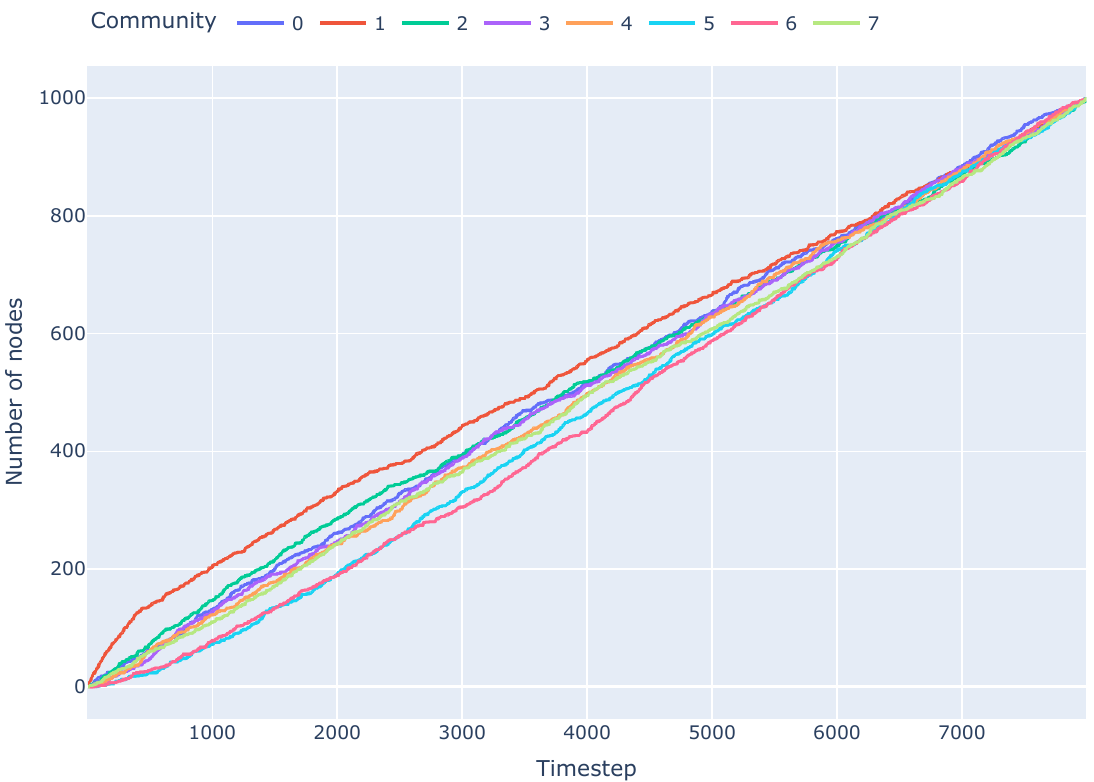}}}
    \caption{Community size evolution}
    
\end{subfigure}

\caption{Sampling using RO (Random Outsider) for a network of 8 blocks with 1000 nodes and one seed node each. The ratio of intra cluster to inter cluster edges ($r$) and average degree within the block ($\langle k' \rangle$) are set as 4 and 10 respectively}
 \label{fig:sbm_1k_1_4_comm_evol_boundary_random}
\end{figure}

In stark contrast, when we investigate one of the random sampling methods, specifically Random Outsider (RO), applied to the identical SBM configuration, Figure \ref{fig:sbm_1k_1_4_comm_evol_boundary_random}(a) conspicuously lacks any discernible inflection points. Likewise, in Figure \ref{fig:sbm_1k_1_4_comm_evol_boundary_random}(b), as anticipated, we notice that the sizes of all communities are simultaneously increasing. This observation indicates that the sampled network does not exhibit a preference for obtaining one community at a time and does not account for community partitions. Similar behavior was observed with the other two random schemes, namely RI\_RO and RI\_MAS.

Having established our desired outcomes from the sampling scheme, we will now explore how its behavior varies across different configurations and assess the limits of detectability for inflection points by varying the values of $r$.

\paragraph{Selection and distribution of seeds.}
Throughout our experimentation, we observed that varying the choice of seed nodes had little to no discernible impact on sampling behavior. In other words, the sampling behavior remained largely consistent whether we opted for higher-degree, randomly selected, or lower-degree nodes. However, it is important to note that sampling does indeed depend on the distribution of seeds across blocks. This phenomenon can be attributed to the behavior of MAS, which tends to greedily favor the nodes with larger boundaries in an effort to minimize the boundary of the cluster. 

\begin{figure}[h!]
\centering
\begin{subfigure}[b]{\linewidth}
    {{\includegraphics[width=\textwidth]{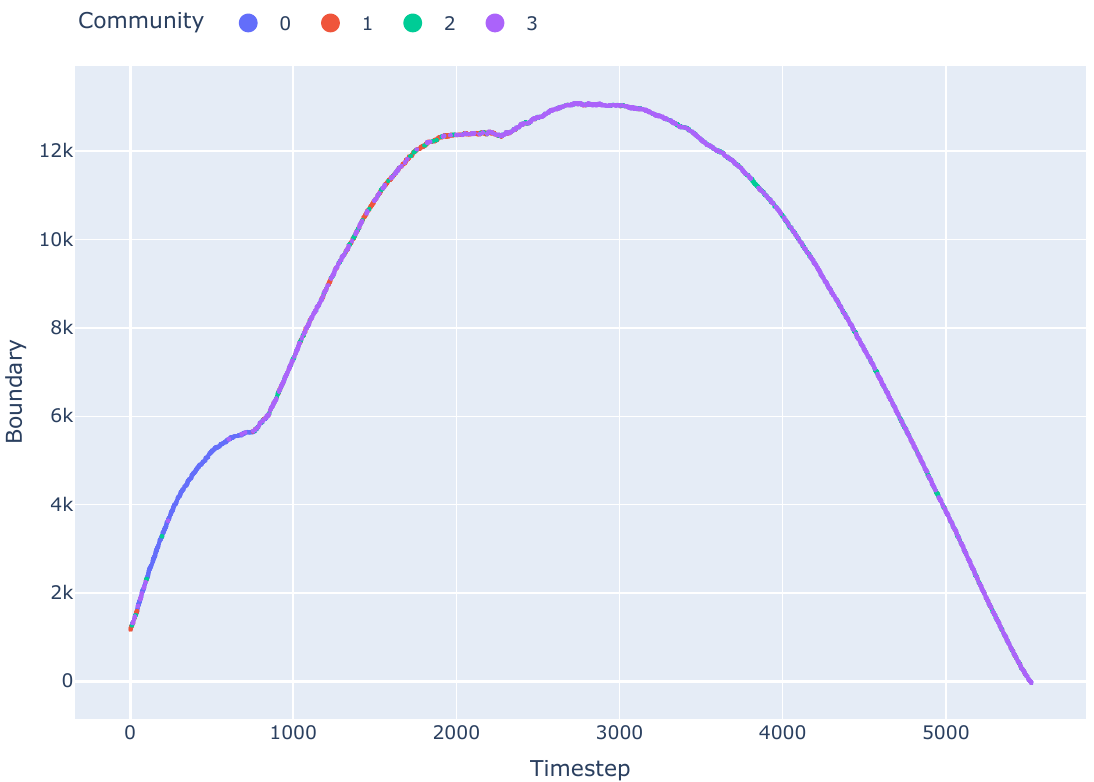}}}
    \caption{Evolution of network boundary with time}
\end{subfigure}
\begin{subfigure}[b]{\linewidth}
    {{\includegraphics[width=\textwidth]{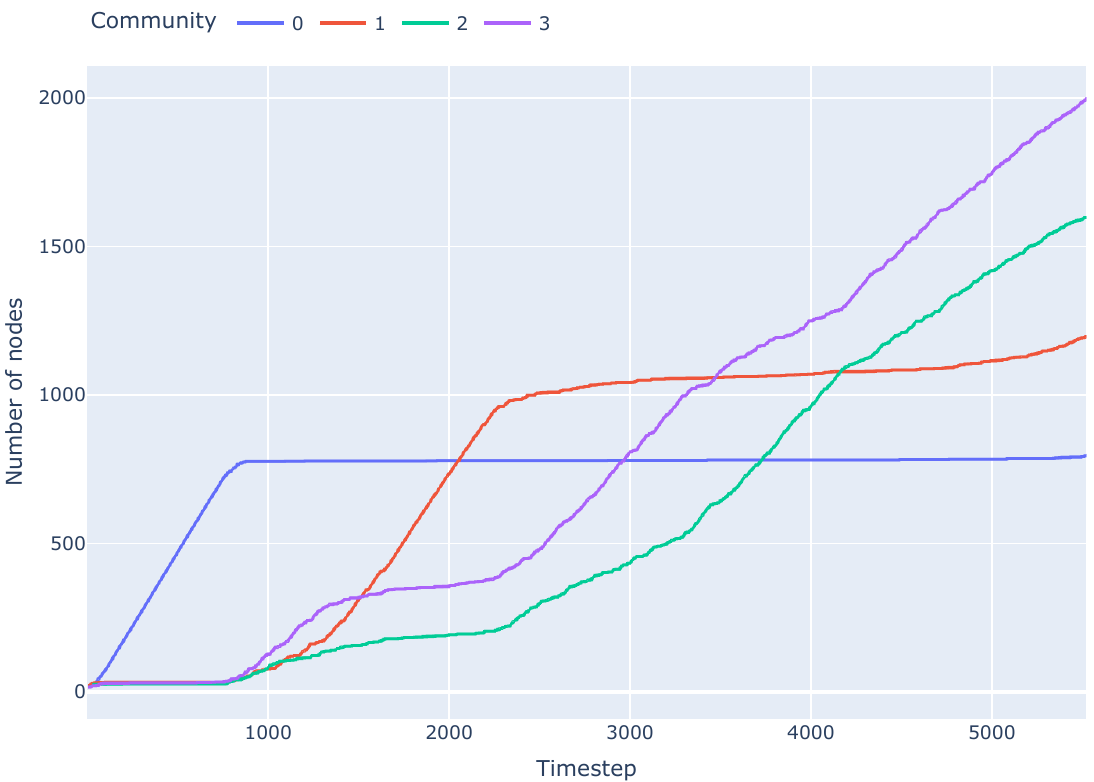}}}
    \caption{Community size evolution}
    
\end{subfigure}

\caption{Sampling using MAS for a network of block sizes \{800, 1200, 1600, 2000\} with 20 seed nodes per block. The ratio of intra cluster to inter cluster edges ($r$) and average degree within the block ($\langle k' \rangle$) are set as 1 and 10 respectively}
 \label{fig:sbm_edge_case}
\end{figure}

\paragraph{Ratio of intra- to inter-cluster edges ($r$).}
In this study, we employ the ratio $r$ as a metric to gauge the 'cohesiveness' of a community. A significant contrast in sampling behavior becomes apparent when $r \geq 2$, leading to the identification of inflection points signifying the transition from one community to another. Conversely, when $r=1$, the inflection points on the boundary vs. timestep plot are not easily discernible, particularly in scenarios where seeds are distributed across multiple blocks of the SBM with differing block sizes. This phenomenon is exemplified in Figure \ref{fig:sbm_edge_case}(a) for the boundary vs. timestep plot of a network characterized by block sizes {800, 1200, 1600, 2000}, $r=1$, and $20$ seed nodes per block.

As depicted in \ref{fig:sbm_edge_case}(b), following the sampling of community `0', we observe that community `1' attempts to be entirely sampled but becomes contaminated with nodes from communities `2' and `3'. After approximately timestep 3200, no particular community exhibits a clear preference for complete sampling.

Nevertheless, in scenarios where communities are of equal size and seed nodes are uniformly distributed, we can still detect inflection points even when $r=1$. Although less visible than those observed when $r \geq 2$, these inflection points remain detectable. For even lower values of $r$, the inflection points become less pronounced, and it becomes apparent that multiple communities are being sampled simultaneously.

Forecasting the exact sequencing of community sampling subsequently becomes complex, as it is influenced by a multitude of factors, including both intra-cluster edges and the inter-cluster edges between the community that has been sampled and those that remain unsampled, all of which impact the directed boundary of the sampled nodes.

\paragraph{Observations.}
Throughout our experimental investigations, we have discerned that the sampler's behavior is contingent upon the seeds' distribution and the intra- to inter-cluster ratio's value ($r$). However, when seeking to determine which community is being sampled at a given point in time, we have found that the plot depicting boundary vs. timestep tends to yield precise insights. We have observed that once a community is exhausted after sampling, a brief period of competition ensues among candidate communities, contending for the next sampling opportunity. The duration of this phase varies depending on the value of $r$. Specifically, for smaller values of $r$, this phase tends to be protracted, leading to the concurrent sampling of nodes from multiple communities.
Conversely, this competition is relatively shorter for larger values of $r$. Eventually, the community with the highest boundary emerges as the winner, attracting a substantial influx of users who follow its initial lead. A higher value of $r$ (such as $r \geq 2$) closely aligns with this ideal behavior since it results in a better-defined community structure, thereby reducing the likelihood of contention. 

\subsection{Empirical data}
As a case study, we expand a well-curated data set of topically relevant Twitter profiles
by sampling additional profiles that form cohesive communities of engagement with them.

The selection of seed profiles is from the DISMISS dataset~\cite{arya_dismiss_2022}, comprising a cohort of $11,580$ highly networked individuals. Since are looking for individuals engaging with information sources, we want our seed set to consist of influential profiles triggering engagement. 
DISMISS is seen as an ideal case in the context of the Indian political sphere.
For our study we focus on a subset of these individuals as seed users, namely those with the `category' label as `civil society'. Here, `category' indicates the `primary industry' of the respective user and can have values like `civil society', `creative', etc.

To facilitate the study, we collected tweets posted by the seed users during July 2022 and use interactions received by these tweets to form a seed network as discussed in the previous section. During the process, we further filter out users to keep only those who posted at least one tweet in the said duration, resulting in $1{,}095$ users as the seed set, from the initial $1{,}184$ belonging to the `civil society' category.

For the above $1{,}095$ seed users, we obtained $50{,}379$ tweets for the chosen duration. To ensure a balanced dataset and mitigate the potential influence of outlier tweets that may have garnered an exceptionally high number of interactions, we employed a ranking approach, focusing on the lower $90\%$ of the tweet interactions. This curation process ultimately yielded a final set of $45{,}341$ tweets, which had received interactions from $379{,}514$ distinct Twitter users. Among the seed users, the number of authored tweets ranged from $1$ to $534$. This curated dataset forms the cornerstone for constructing an initial network, which subsequently serves as the foundational point for ongoing data collection efforts pertaining to Distinct, Nested, and A-F sampling schemes. The data collection was initiated in December 2022 under the presumption that interactions had reached a stable state by that time.

The interactions from these collected tweets were used to get weights per interaction type for all the three schemes - distinct, nested and audience-facing, as shown in Table \ref{tab:weights}.

\subsubsection{Sampling:} 
For the seed network generated above, along with the weights from Table \ref{tab:weights}, we sample the networks for distinct, nested and audience-facing variants.
Along with the three variants, we also sample using four types of random sampling schemes for comparison.

Our random node sampling strategy possesses two key attributes: selection probabilities and selection strategy. Selection probabilities can either be uniform ($U$), where all samples share an equal probability of being chosen, or weighted ($W$), where the probability is determined based on the priority/score computed using our sampling scheme.
The selection strategy encompasses two options: `direct' ($D$), in which one of the outsiders is chosen randomly with the specified selection probability, and `staged' ($S$), which involves the selection of an insider at random, followed by the selection of one of the chosen insider's outsiders with the given selection probability.
By combining these features, we derive four distinct random node sampling strategies, denoted as $RS\_DU$, $RS\_DW$, $RS\_SU$, and $RS\_SW$ by considering all possible combinations. Table~\ref{tab:dataset-stats} provides an overview of the Twitter data sampled, using the three variants of our sampling scheme in conjunction with the random node-based sampling strategies. It is worth noting that due to the Twitter API shutdown, the sizes of the collected sampled networks differ, ranging from a minimum of $1{,}905$  for $RS\_DW$ to a maximum of $5{,}515$ for $RS\_SU$. 

\begin{figure*}[!ht] 
\centering 
\subfloat[\centering ][Community evolution of 8 largest communities using A-F]
{{\includegraphics[width=0.45\textwidth]{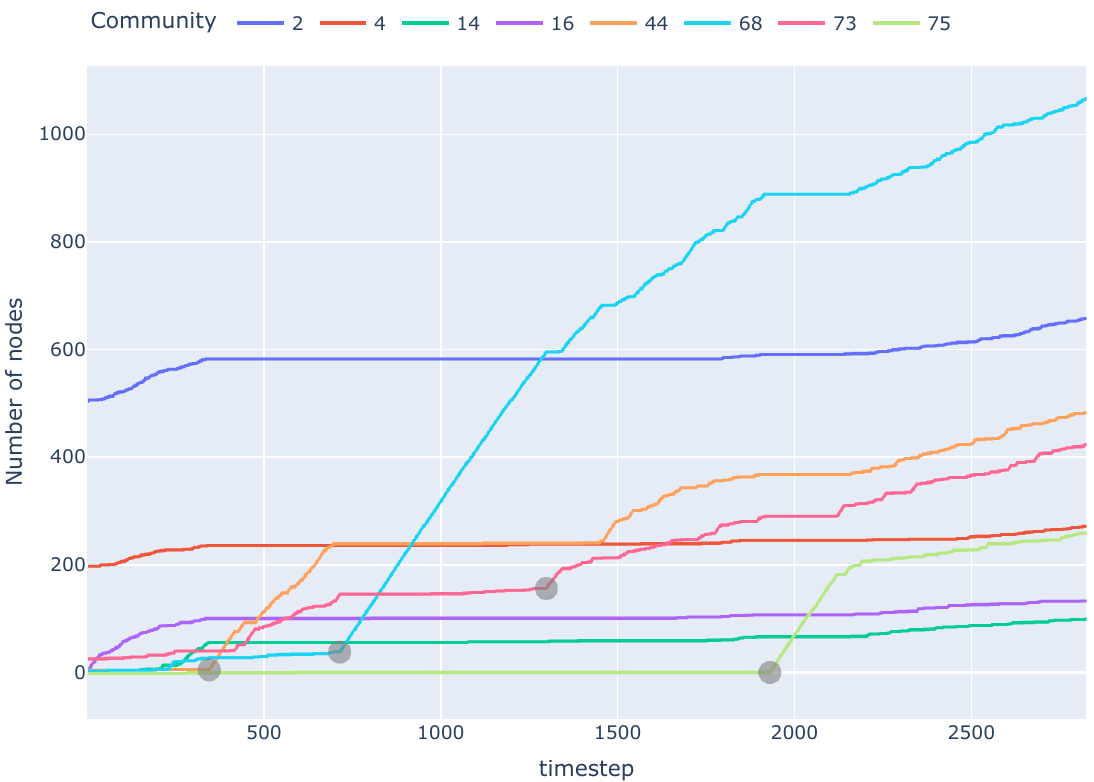}}}
\subfloat[\centering ][Change in boundary using A-F]
{{\includegraphics[width=0.45\textwidth]{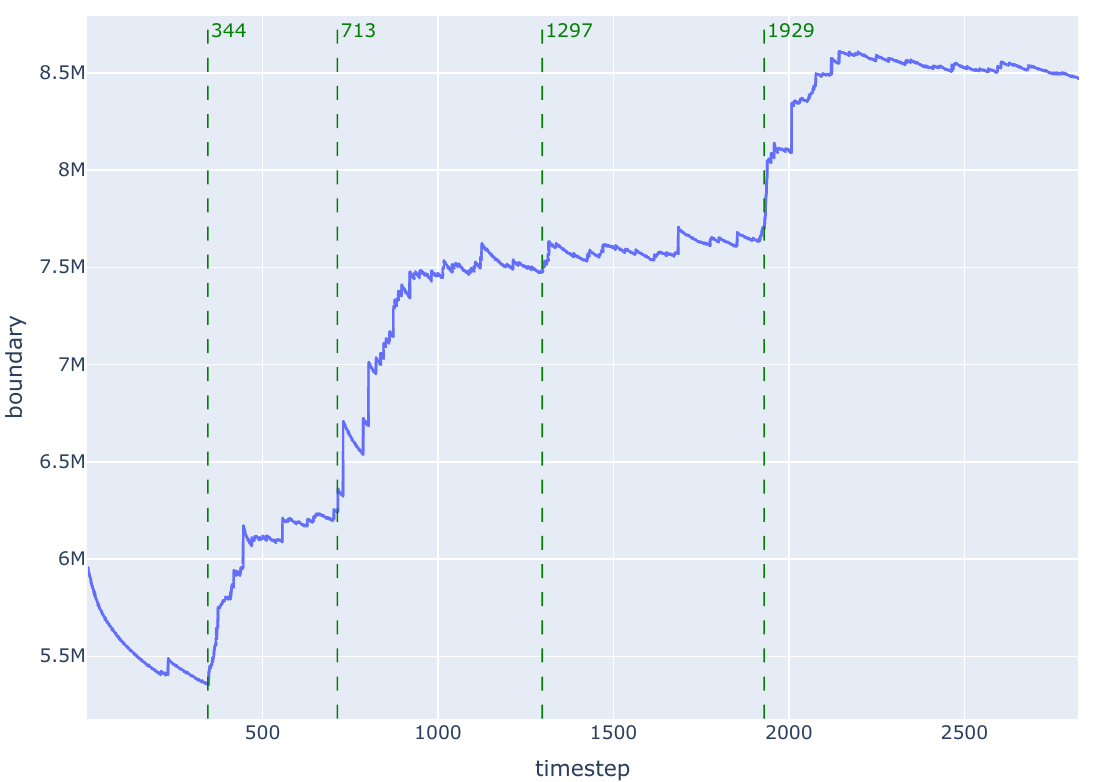}}}
\caption{The four shaded circles in (a) indicate points where significant sampling of a new community starts. For corresponding timesteps in (b), we observe that boundary shoots up before plateauing again. This is especially noticeable for timesteps = \{344, 713, 1929\}. For a smaller increase in community size, as seen for timestep = 1929, we still also observe a small rise in boundary value in (b). However, not every small rise in (b) corresponds to a different community being sampled in (a). Hence, we tend to focus on bigger jumps in value of boundary timestep = \{344, 713, 1929\} where a different community begins sampling as can be seen through the steep rise in community size.}
 \label{fig:tab_comm_evol_boundary}
\end{figure*}


    


\begin{table}
\centerline{\footnotesize
\begin{tabular}{|l|r|r|r|r|r|}
\hline
\textbf{Sampling} & \textbf{In-}& & \textbf{Edges in} & \textbf{Total} & \\
\textbf{Scheme} & \textbf{siders} & \textbf{Nodes} & \textbf{insider} & \textbf{edges} & \textbf{Tweets} \\
 &  & & \textbf{network} &  & \\ \hline
\textbf{Distinct}   & 8,721 & 609,609 & 208,628 & 1,545,420 & 161,471 \\
\textbf{Nested}   & 4,698 & 525,531 & 98,889  & 1,182,774 & 91,966  \\
\textbf{A-F}   & 3,919 & 513,466 & 93,476  & 1,149,281 & 84,267  \\
\textbf{RS\_DU} & 1,976 & 417,439 & 5,438   & 745,871   & 50,856  \\
\textbf{RS\_DW} & 1,905 & 410,061 & 8,383   & 744,067   & 51,191  \\
\textbf{RS\_SU} & 5,515 & 600,858 & 28,536  & 1,070,803 & 74,463  \\
\textbf{RS\_SW} & 3,355 & 527,265 & 34,127  & 1,023,682 & 62,872  \\ \hline
\end{tabular}}
\caption{Dataset statistics for the sampled Twitter network using the three variants of our sampling scheme and four variants of random sampling.}
\label{tab:dataset-stats} 
\end{table}

\begin{table*}[ht!]
\centering
\begin{tabular}{|ll|l|l|l|l|}
\hline
\multicolumn{2}{|l|}{\textbf{Sampling scheme}}                                         & $CC_{local}$ & $CC_{global}$ & $\langle L \rangle$    & $\langle k \rangle$     \\ \hline
\multicolumn{1}{|l|}{\multirow{3}{*}{Priority}} & Distinct                    & 0.2566     & \textbf{0.4239}      & 5.34 & 12.97 \\
\multicolumn{1}{|l|}{}                          & Nested                      & 0.3747     & 0.4145      & 4.62 & 21.65 \\
\multicolumn{1}{|l|}{}                          & Audience-Facing (A-F) & \textbf{0.4004}     & 0.4035      & \textbf{4.40} & \textbf{26.49} \\ \hline
\multicolumn{1}{|l|}{\multirow{4}{*}{Random}}   & RS\_DU                      & 0.0646     & 0.0698      & 5.25 & 3.40  \\
\multicolumn{1}{|l|}{}                          & RS\_DW                      & 0.1360     & 0.0608      & 4.87 & 5.32  \\
\multicolumn{1}{|l|}{}                          & RS\_SU                      & 0.1179     & 0.0559      & 4.95 & 4.81  \\
\multicolumn{1}{|l|}{}                          & RS\_SW                      & 0.1237     & 0.0562      & 4.33 & 9.11  \\ \hline
\end{tabular}
\caption{Sampling schemes (Distinct, nested, and audience-facing) evaluated on Twitter. Here, $CC_{local}$, $CC_{global}$, $\langle L \rangle$, $\langle k \rangle$ refer to the average of local clustering coefficients, global clustering coefficient, average shortest path, and average degree respectively. The bold values signify the highest or lowest values as per the chosen metric. To ensure comparison across sampling schemes despite different sampled network sizes, we consider the subgraphs sampled till the minimum common network size and calculate metrics for that snapshot. Using this approach, we have one common timestep for the priority based sampling schemes and one for random sampling schemes.}
\label{tab:metrics-v3}
\end{table*}

\subsubsection{Evaluation:}
The fundamental distinction between the Twitter network we collected and the synthetic networks we generated pertains to the definition of a community. In the case of synthetic networks, a community was explicitly defined as a block utilized in configuring the Stochastic Block Model (SBM). In contrast, with the Twitter network, we lack a definitive ``community label" and must rely on obtaining it without a guarantee of accuracy. In an effort to potentially assign community labels to each node, we apply the Louvain community detection algorithm to the collected Twitter network. Analogous to our approach with synthetic networks, we employ these community labels to explore potential correlations between the initiation of community sampling and the boundary vs. timestep plot of the entire network.

As observed in synthetic networks, we anticipate the presence of inflection points in the ``boundary vs timestep" plots, indicating the transition from sampling one community to the next. In the context of the Twitter network, we notice a similar pattern, although occasional instances occur where two communities are sampled concurrently. For instance, in the case of sampling using the Audience-Facing (A-F) approach, exemplified in Figure~\ref{fig:tab_comm_evol_boundary}(a), we discern distinct segments where only one community is sampled at any given time. However, there are intervals during which, alongside the primary community, certain nodes from a background community are also included in the insider set. This occurrence is linked to scenarios where the priority of nodes is identical, signifying they possess similar weighted directed boundary values. A notable example of this behavior can be found in the time range between timesteps $t=800$ and $t=1200$, where we observe substantial growth in the community labeled as ``68" (blue), while a few nodes are added to the community labeled as ``73" (pink).

Despite the utilization of community labels generated from the data, we are still able to identify significant spikes in boundary values that correspond to the initiation of new communities. In Figure \ref{fig:tab_comm_evol_boundary}(b), we can observe these spikes corresponding to the commencement of communities ``44," ``68," and ``75" at timesteps 344, 713, and 1929, respectively. Following this initiation, the boundary values stabilize briefly before witnessing another spike with the onset of a new community. It is worth highlighting that at timestep $1297$, where although the expansion of community ``73" is relatively modest, the boundary plot effectively captures it with a minor peak. In a real-world scenario, however, users would encounter a boundary-versus-timestep plot and need to discern points such as the one at timestep $1297$, which might be easily overlooked amidst the noise. Nevertheless, the approach remains capable of capturing rising trends akin to those observed at timesteps \{$344$, $713$, $1929$\}.

\paragraph{Structure based metrics}
To gain insight into the characteristics of the cohesive communities obtained through sampling, we conduct a comprehensive evaluation using informative structural metrics, as outlined below:

\begin{enumerate}
    \item \emph{Average shortest path ($\langle L \rangle$):} The average directed  path length along the shortest paths for all possible pairs of nodes. 
    \item \emph{Clustering Coefficient:} 
        \begin{enumerate}
            \item \emph{Local clustering coefficient ($CC_{local}$)}: The  local clustering coefficient for a node $i$ on a directed network is given by

            \begin{equation*}
                CC_i = \frac{|{e_{jk} : j,k \in N(i), \;e_{jk} \in E}|}{deg(i)(deg(i)-1)}
            \end{equation*}
            where $E$ denotes the set of edges in the graph and $N(i)$ denotes the open neighborhood of node $i$. The average local clustering coefficient, $CC_{local}$, is the mean of the local clustering coefficients of all nodes.
            \item \emph{Global clustering coefficient ($CC_{global}$)}: This is given by the ratio of the number of closed triplets over all possible triplets in the network.
        \end{enumerate}
    \item \emph{Average degree ($\langle k \rangle$):} The average degree is the mean of all node degrees. 
\end{enumerate}

As the sizes of the sampled networks obtained through the three variants and four random schemes vary, we restrict our analysis to the subgraphs sampled up to the size of the smallest common network. This approach allows us to calculate metrics on networks of equal size, ensuring the comparability of results across different schemes while eliminating the influence of network size disparities.

As presented in Table \ref{tab:metrics-v3}, we observe that the clustering values ($CC_{local}$ and $CC_{global}$) for the proposed priority-based sampling schemes are notably higher compared to any of the random sampling schemes. This disparity in values suggests that networks obtained through priority-based schemes exhibit stronger connectivity. Additionally, Table \ref{tab:metrics-v3} reveals that the Audience Facing interactions variant outperforms all other variants in terms of $CC_{local}$ metric, indicating a higher number of triads, with the exception of $CC_{global}$ where the distinct variant maintains a slight advantage. It is also important to highlight that all the variants have a significant performance advantage over any random sampling schemes.

\section{Conclusion and Future Work}
We proposed a novel scheme for snowball-type sampling in unbounded networks 
designed to respect cohesive communities.
Its intended purpose is the extraction of communities. 

Our approach consists of two main parts, 
a sampling priority utilizing the maximum-adjacency principle,
and a method to integrate modes of interaction into a single weighted directed graph.
The latter is based on importance scaling and can be calibrated empirically
as demonstrated in a prototypical case study.
Computational experiments on synthetic and empirical data demonstrate
that our method samples subgraphs with low inwards-directed conductance 
by keeping the boundary around the sampled region small.
While the growth inside communities is 
almost perfect in the idealized setting of stochastic blockmodels,
a similar evolution is observed in the case study on Twitter
that motivated this research.

With an adapted construction of the weighted graph,
our sampling strategy transfers to other social media networks such as Reddit or Facebook,
and it will be interesting to apply the maximum-adjacency principle in other settings
such as the respondent-driven sampling of offline social networks \cite{heckathorn1997rds}.

\section{Ethics Statement}
In this work, we use Twitter API to gather a stream of tweets and their interactions. 
We leverage only publicly available information at the time of data collection. 
To comply with the terms of service, we will release our anonymized data (network structure) for research purposes only.
To the best of our knowledge, no code of ethics has been violated throughout the data collection and experimental phase in this study.

\bibliography{files/bibfile}

\end{document}